\documentclass[letterpaper,twocolumn,english,aps,prb,groupedaddress,superscriptaddress,showpacs,amsfonts]{revtex4}
\usepackage{ae}
\usepackage{aecompl}
\usepackage[T1]{fontenc}
\usepackage[latin1]{inputenc}
\usepackage{float}
\usepackage{amsmath}
\usepackage{graphicx}
\usepackage{amssymb}

\makeatletter
\usepackage{babel}
\makeatother
\begin{document}

\title{Microscopic Theory of the Single Impurity Surface Kondo Resonance}

\author{Chiung-Yuan Lin}

\affiliation{Department of Physics, Boston University, 590 Commonwealth Ave.,
Boston, MA 02215}
 
\author{A.~H. Castro Neto}

\affiliation{Department of Physics, Boston University,
590 Commonwealth Ave., Boston, MA 02215}

\author{B.~A.~Jones}

\affiliation{IBM Almaden Research Center, San Jose, CA 95120-6099}

\date{\today}

\begin{abstract}
We develop a microscopic theory of the single impurity Kondo effect
on a metallic surface. We calculate the hybridization
energies for the Anderson Hamiltonian of a magnetic impurity interacting
with surface and bulk states and show that, contrary to the Kondo effect
of an impurity in the bulk,
the hybridization matrix elements are strongly dependent on the momentum
around the Fermi surface. Furthermore, by calculating the tunneling conductance
of a scanning tunneling microscope (STM), we show that when the magnetic
impurity is located at a surface the Kondo effect can occur with equal
strength between bulk and surface states. We compare our results with
recent experiments of Co impurities in Cu(111) and Cu(100) surfaces
and find good quantitative agreement.
\end{abstract}

\pacs{72.15.Qm,68.37.Ef,72.10.Fk}

\maketitle

\section{Introduction}

When a magnetic impurity is located in
the bulk of a metal it undergoes a non-trivial many-body scattering
with the  electronic states at the Fermi energy, $\epsilon_F$, called the Kondo
effect \cite{kondo}.
The bulk Kondo effect (to be contrasted with the surface Kondo
effect studied in this paper) is one of the best
studied problems in condensed matter physics and many different
techniques from renormalization group \cite{wilson} to Bethe
ansatz \cite{bethe} have been used over the years.
The basic mechanism of the Kondo effect is the magnetic screening
of the impurity (the formation of the Kondo singlet)
at temperatures $T$ below the Kondo temperature $T_K$. Above $T_K$
the magnetic impurity behaves paramagnetically but for $T<T_K$
a resonance is formed close to the Fermi surface.
The Kondo effect is very important in modern condensed matter theory
and appears in many different areas of research, from U and Ce
intermetallics (heavy-fermions) \cite{heavy_fermion} to quantum
dots \cite{quantum_dots}. There is a wide variety of phenomena
that can be described within the universality of the Kondo problem.
The Kondo effect can be observed experimentally in measurements of
the temperature dependence of resistivity (the so-called resistivity
minimum) \cite{resistivity}, and also in thermodynamic measurements
such as specific heat and magnetic susceptibility \cite{heavy_fermion}
of dilute magnetic alloys, due to the enhancement of the density
of states close to the Fermi energy (the Abrikosov-Suhl
resonance \cite{abrikosov_suhl}). The enhancement of the density
of states is related to the change of the characteristic energy
scales from $\epsilon_F$ to $k_B T_K$ and therefore in the density of states
from $N(0) \propto 1/\epsilon_F$ to $N^*(0)\propto 1/(k_B T_K) \gg N(0)$
since $\epsilon_F \gg k_B T_K$.

\begin{figure}[htb]
\begin{center}
\includegraphics[width=6cm,keepaspectratio]{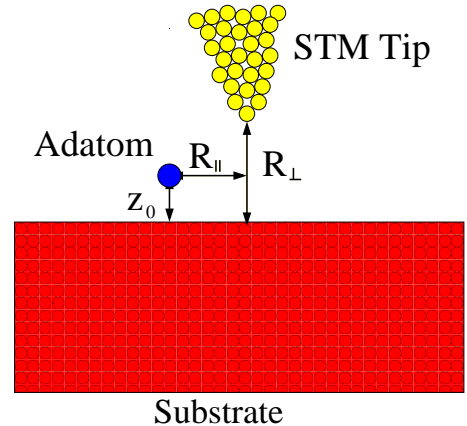}
\end{center}
\caption{Schematic picture of a STM measurement: $z_0$ is the distance
of the adatom to the surface, $R_{\perp}$ is the distance of the STM
tip from the surface, and $R_{||}$ is the distance, along the surface,
between the STM tip and the adatom.}
\label{fig-system}
\end{figure}

The Kondo effect was also observed recently in STM studies of
magnetic atoms on metallic surfaces \cite{Jintao,Crommie,Eigler}
(see Fig.\ref{fig-system}).
In a STM experiment electrons from a sharp tip tunnel into
the material to be studied, creating a tunneling current $I$ due
to the application of a potential $V$.
Roughly speaking a STM experiment measures the local density
of states at the Fermi energy via the
differential tunneling conductance $dI/dV$ \cite{Gadzuk}. 
When a STM tip is
away from the magnetic impurity it measures
the substrate density of states, $N(0)$; however, close
to the impurity (that we call the adatom)
electrons from the tip can tunnel directly
to the impurity. The theory
of STM is far from trivial because electrons
from the tip not only tunnel to the adatom but also
to the bulk and surface states, that is, there
are various different channels of tunneling that lead to
interference effects that have to be taken
into consideration for the proper interpretation
of the experimental data.

The basic starting point for the study of the Kondo effect is
the Anderson impurity Hamiltonian \cite{anderson}:
\begin{eqnarray}
H_{s}&=& \sum_a \left(\epsilon_{a}\sum_{\sigma}c_{a\sigma}^{\dagger}c_{a\sigma}
+U n_{a\uparrow}n_{a\downarrow}\right)
+\sum_{{\bf k}\sigma}\epsilon_{k}c_{{\bf k}\sigma}^{\dagger}c_{{\bf k}\sigma}
\nonumber \\
&&+\sum_{{\bf k},\sigma,a}t_{{\bf k}a}\left(c_{{\bf k}\sigma}^{\dagger}c_{a\sigma}
+{\rm{H}}.{\rm{c}}.\right), \label{Hamiltonian}
\end{eqnarray}
where $c_{a\sigma}$ ($c^{\dagger}_{a\sigma}$) is the annihilation
(creation) operator for an electron on a localized atomic state
(impurity state) with angular momentum labelled by $a$, 
energy $\epsilon_{a}$ and spin $\sigma$
($\sigma=\uparrow,\downarrow$),
$c_{{\bf k}\sigma}$ ($c_{{\bf k}\sigma}^{\dagger}$) is the
annihilation (creation) operator for an electron on the conduction
band with dispersion $\epsilon_{k}$ and momentum ${\bf k}$,
$n_{a\sigma} = c^{\dagger}_{a\sigma} c_{a\sigma}$ is the number
operator, $U$ is the Coulomb energy for the double occupancy of
the impurity state, and $t_{{\bf k}a}$ is the
hybridization energy between the impurity and conduction
states (the electron operators obey anti-commutation rules:
$\{c_a,c^{\dagger}_b\} = \delta_{a,b}$).

One of the characteristics of the Anderson impurity model is the
distinction between substrate and adatom wavefunctions. Although
most theoretical works
do not question the distinction made {\it a priori} between
these quantum states, it turns out that this distinction 
is not completely natural. 
The reason is very simple: when an impurity atom is introduced in a
metal, it hybridizes with the metallic states losing its identity.
However, it
leaves behind a phase shift $\delta_{{\bf k}}$ in the original 
metallic states. 
Thus, the impurity state cannot be really distinguished from the
host states from the quantum mechanical point of view. Nevertheless,
Anderson \cite{anderson} makes the point that because the d-orbitals 
are a inner shell, the Coulomb energy $U$ for double
occupancy of those orbitals is large and they must be distinguished
from metallic states where the Coulomb energy is strongly suppressed
by screening effects. Thus, the distinction between these two types
of states can only be clearly made when these states are orthogonal
to each other so that the impurity does not cause a direct perturbation
in the substrate spectrum. Nevertheless, even in
metals where the electronic bands are generated out of d-orbitals (as
in the case of Cu), the strong metallic bonding leads to a large s-wave
character of the bands and to very small overlap with the adatom d-orbitals
(that is, these states are ``naturally'' orthogonal \cite{kfr}). 
Nevertheless, this orthogonality can only be distinguished {\it a posteriori}.
In fact, we have found, by direct numerical computation, that this is 
the case in the surface Kondo effect.
Finally, as pointed out by Anderson \cite{anderson} 
(see also ref.\cite{tsvelik}),
the orthogonality of these states is not fundamental for the magnetic
phenomenon which is essentially a local effect and all the subtleties
associated with orthogonality become encapsulated into the hybridization
matrix elements $t_{{\bf k}a}$ which become a phenomenological parameter to be
obtained indirectly from the experiment. However, in trying to understand
the STM experiments, and especially the
role played by the surface and bulk states, we cannot simply take these
matrix elements as phenomenological parameters since we would not be able
to separate the contributions coming from the bulk and the surface of the
Cu substrate.

Hence, in this paper we follow the scheme proposed by Tsvelik and
Wiegmann \cite{tsvelik} and work with non-orthogonalized states.
In this case the hybridization energies are given by: 
\begin{eqnarray}
t_{{\bf k}a} = (\epsilon_k + \epsilon_a) \int d^3 r \, \, \psi^*_a({\bf r})
\psi_{{\bf k}}({\bf r})
\end{eqnarray}
where $\psi_a({\bf r})$ is the atomic state of the impurity which is localized
on the impurity site (that is, it decays exponentially away from the impurity)
and $\psi_{{\bf k}}({\bf r})$ is the Bloch wavefunction of the crystal.
The hybridization energy depends not only on the amplitude
but also on the direction of ${\bf k}$. Nevertheless, 
it is usually assumed that the hybridization is constant in momentum
space, that is, $t_{{\bf k},a} = t$.
Although this assumption maybe warranted for spherical gapless Fermi
surfaces (since the momentum dependence ``averages out" and most of the interesting
physics occurs at the Fermi surface) it is not valid for metals where the
chemical potential crosses gaps in the electronic spectrum. 
The Kondo Hamiltonian can be obtained from (\ref{Hamiltonian})
in the limit of $|t| \ll \epsilon_a, U$ via the Schrieffer-Wolf
transformation \cite{schrieffer_wolf} that maps the Hamiltonian
(\ref{Hamiltonian}) onto:
\begin{eqnarray}
 H_{K}= \sum_{{\bf k}\sigma}\epsilon_{k}c_{{\bf k}\sigma}^{\dagger}c_{{\bf k}\sigma}
+ J_K \,  {\bf S} \cdot {\bf s}(0)
\label{Kondo_Hamiltonian_1}
\end{eqnarray}
where ${\bf S}$ is the impurity spin,
$s^a(0) = \hbar/2  \sum_{{\bf k},{\bf k'},\alpha,\beta}
c_{{\bf k}\alpha}^{\dagger} \sigma^a_{\alpha,\beta} c_{{\bf k'}\beta}$
is the electron spin at the impurity site ($\sigma^a$ with $a=x,y,z$ are Pauli
matrices) and
\begin{eqnarray}
J_K \approx |t|^2 \left(\frac{1}{|\epsilon_a|}+\frac{1}{|\epsilon_a+U|}\right)
\end{eqnarray}
is the exchange interaction between local spins and conduction electrons (the
energy is measured relative to the Fermi energy).
In going from (\ref{Hamiltonian}) to (\ref{Kondo_Hamiltonian_1}) we have
disregarded a series of physical processes that cannot be described as
spin exchange interactions, foremost being the
variable occupancy of the impurity site.

\begin{figure}[htb]
\begin{center}
\includegraphics[width=6cm,keepaspectratio]{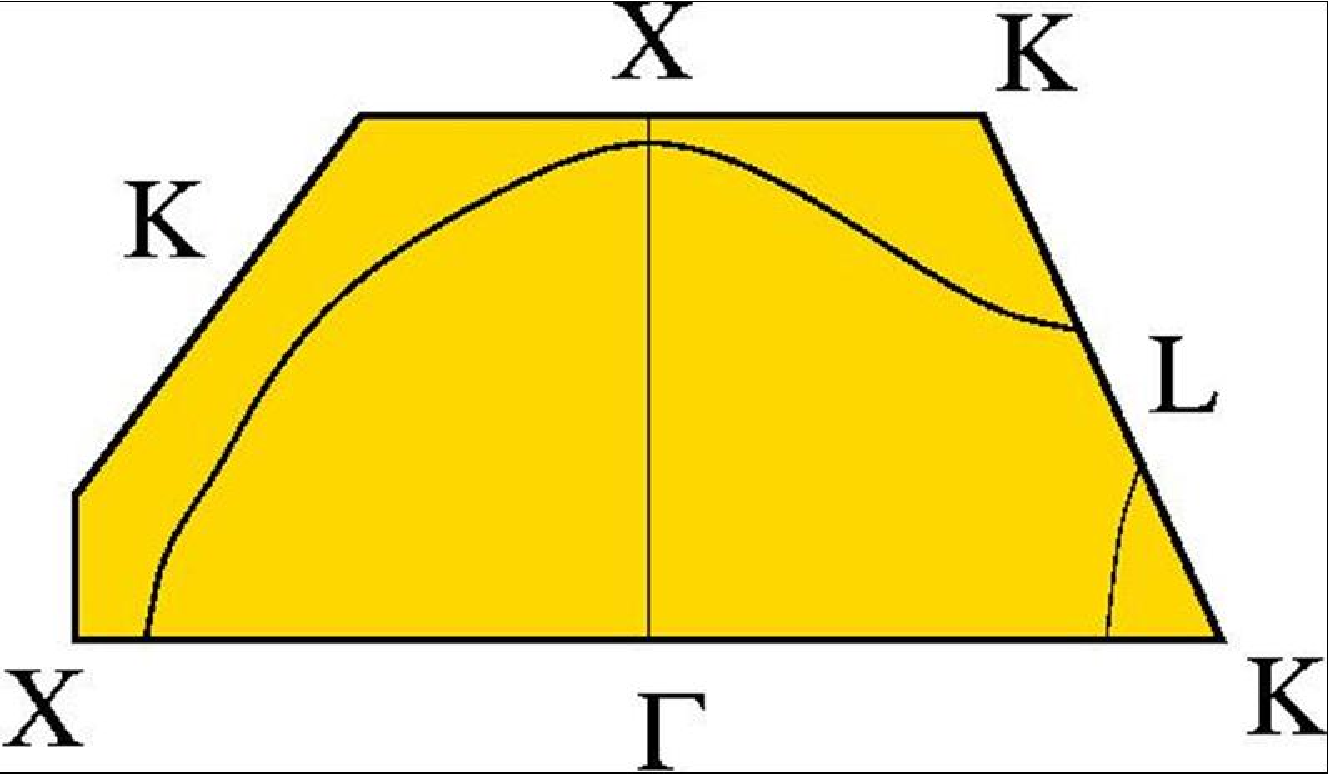}
\end{center}
\caption{Planar cut of the bulk Cu Fermi surface in the first Brillouin
zone: $\Gamma$-L is the
(111) direction (notice the absence of states due to the bulk gap);
$\Gamma$-X is the (100) direction.}
\label{fs}
\end{figure}

Even simple metals like Cu have energy gaps in their electronic
spectrum due to the periodic potential generated by the Cu ions
\cite{paper_cu,Knapp}. For instance, it is well-known that for Cu along
the (111) direction there is a gap of $5.1$ eV at the Fermi
energy while there are no gaps along the (100) direction (see Fig.\ref{fs}). 
On the one hand, $t_{{\bf k}a}$ does not exist for bulk states when ${\bf
k}$ points along the (111) direction. However, because of the bulk
gap, one has a surface state in the Cu(111) surface
\cite{surface_physics} (a state that decays exponentially in the
(111) direction but it is extended in the directions perpendicular
to (111)) that can hybridize with the impurity located on the
Cu(111) surface. On the other hand, there are no surface states in the 
(100) direction to hybridize with an impurity located on the Cu(100)
surface, although bulk states hybridize with it since their
wavefunctions decay exponentially outside of the crystal. 
Previous studies \cite{Ujsaghy} on the surface Kondo effect either assume a
momentum-independent hybridization $t_{ka}$ or sum
average over a gapless Fermi surface \cite{Gadzuk}. By ignoring
momentum dependence or the gap structure of the Fermi surface, the
comparison between Cu(111) and Cu(100) surface Kondo effects of noble
metals is not possible. In fact, we show that recent
STM experiments of atoms located in the Cu(111) and Cu(100) surfaces
have been interpreted incorrectly because of these differences. In
particular, we show that the surface Kondo effect between a
magnetic atom located on the Cu(111) surface and the Cu(111) surface state
is quantitatively similar to the surface Kondo effect of a
magnetic atom on a Cu(100) surface with bulk states. The main reason
for this similarity comes from the fact that the bulk state that
dominates the Kondo scattering is the one with the largest
penetration in the work-function barrier, that is, the state with
momentum perpendicular to the Cu(100) surface. The state 
with momentum perpendicular to the Cu(100) surface has
essentially the same decay rate out of the crystal as the surface
state in the Cu(111) surface and therefore similar hybridization
with the magnetic impurity.

In the presence of the tip new terms
have to be added to the Hamiltonian (1) that describe the
tunneling processes of tip-to-adatom and tip-to-substrate \cite{Gadzuk}.
The tunneling processes of tip-to-adatom is given by:
\begin{equation}
H_{at}=\sum_{\sigma,a}t_{ap}\left(c_{a\sigma}^{\dagger}c_{p\sigma}
+{\rm{H}}.{\rm{c}}.\right).
\end{equation}
where $c_{p\sigma}$ ($c^{\dagger}_{p\sigma}$) annihilates (creates)
and electron at the tip and $t_{ap}$ is the hybridization energy
between tip and adatom. The tip-to-substrate hybridization is given
by:
\begin{equation}
H_{t}=\sum_{k\sigma}t_{{\bf k}p}\left(c_{{\bf k}\sigma}^{\dagger}c_{p\sigma}
+{\rm{H}}.{\rm{c}}.\right),
\end{equation}
where $t_{{\bf k}p}$ is the hybridization energy between substrate (bulk or
surface state) and adatom. The tip Hamiltonian is simply
\begin{equation}
H_{st}=\sum_{\sigma}\epsilon_p c_{p\sigma}^{\dagger}c_{p\sigma} \,
\end{equation}
where $\epsilon_p$ is the energy of the electron states on the tip.
The total Hamiltonian of the tip-substrate-adsorbate system is
$H=H_{s}+H_{t}+H_{at}+H_{st}$. As in the case of (\ref{Hamiltonian})
the hybridization energies are given by:
\begin{equation}
t_{\alpha\beta}=(\epsilon_{\alpha}+\epsilon_{\beta})
\int{d}^{3}r\,\psi_{\alpha}^{\ast}({\bf r})\psi_{\beta}({\bf r})
\label{tunneling}
\end{equation}
with $\alpha,\beta={\bf k},a, p$ labeling the wavefunctions,
$\psi_{\alpha}({\bf r})$, for substrate, adatom and tip, respectively.
Notice that the complexity in this problem comes from the interference
between the different channels of tunneling.

In the present work, we will consider the case of Co on Cu
surfaces while our theory is applicable to any magnetic adatom on
noble metal surfaces. Our main objectives are: ({\it i}) the calculation
of the hybridization matrix elements between the different
wavefunctions in the problem (bulk, surface, impurity, and STM
tip) with the smallest number of adjustable parameters, 
and ({\it ii}) 
the description of the surface Kondo effect via STM measurements. 
While the
spectrum of bulk and surface states is readily available in the
literature, very little has been published on the actual form of
the states. Instead of embarking on a complicated calculation of
wavefunctions via heavy numerical techniques, we opted for a simpler 
approach that can
provide quantitative results that can be directly compared to the
experiment as well as intuitive understanding of the problem. As
we are going to show, our simplified models have limitations in
explaining some of the available STM data and more detailed work
is required. We hope our work stimulates other works that make
use of more sophisticated methods to calculate the matrix
elements because, as we are going to show, they are fundamental
for the understanding of magnetism in metallic surfaces. We model
the periodic potential inside of the crystal in the nearly-free
electron approximation and obtain the bulk gaps as observed in
metallic Cu. The surface states are obtained after the modeling of
the surface potential in terms of the image charge in the vacuum.
The energies of surface states and resonances in each surface are
in very good agreement with  photoemission results in Cu(111) and
Cu(100) surfaces. Following the literature in STM theory, the
impurity is modeled via standard Hydrogen-like wave
functions.

The paper is organized as follows: In Sec.~\ref{sec-wavefunction},
we model the ionic potentials that give rise to the wavefunctions
of the metallic host, adatom, and the STM tip. Using the
wavefunctions and their spectrum we discuss the Cu band structure,
both the bulk and Cu(111) surface state. In Sec.~\ref{sec-hybridization} we
present the results for the hybridization matrix elements that
determine the hybridization energies as a function of the distance
from the surface and adatom and also as a function of the
momentum. The surface Kondo effect is discussed in
Sec.~\ref{sec-surface-Kondo} in the context of the Schrieffer-Wolf
transformation and the slave-boson mean-field theory. Using the
calculated hybridization energies the differential conductance is
calculated in Sec.~\ref{sec-STM} and compared with the STM data.
Sec.~\ref{sec-conclusions} contains our conclusions.

\section{Wavefunctions and Spectrum}
\label{sec-wavefunction}

In this section, we obtain wavefunctions needed to
calculate the tunneling matrix elements (\ref{tunneling}). This
is the first step towards the STM differential conductance. As discussed
in the introduction we are going to use simple models for the various
potentials involved in the problem so that we can make analytical progress
in calculating the hybridization energies.

\begin{figure}[htb]
\begin{center}
\includegraphics[width=7cm,keepaspectratio,angle=-90]{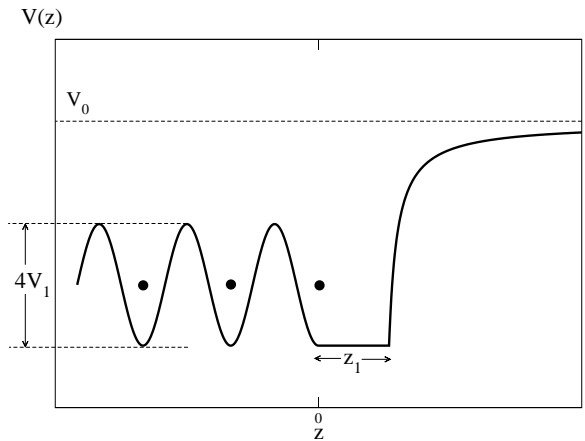}
\end{center}
\caption{Plot of the image potential
energy near a crystal surface. The solid dots are crystal atoms.}
\label{fig-potential}
\end{figure}

Metallic Cu is a very good metal with a nearly spherical Fermi
surface except for the gaps in the (111) direction. These gaps
have their origin in the periodic potential generated by the ions
(see Fig.\ref{fs}).
Although the Cu states can be obtained from first principle
calculations \cite{paper_cu} we adopt the nearly free
electron approximation and model the bulk potential with a simple
periodic function. Furthermore, in order to study the surface
states we consider a semi-infinite crystal where the bulk is
located at $z<0$ (see Fig.~\ref{fig-potential}). We write
the potential in the substrate as:
\begin{eqnarray}
V({\bf r})=\left\{\begin{array}{ll}
-2V_{1}\cos\frac{2\pi z}{a}&\mbox{, for }z<0, \\
-2V_{1}&\mbox{, for }0<z<z_1, \\
V_{0}-\frac{e^{2}}{4(z-z_{\rm im})}&\mbox{, for }z>z_1 ,
\end{array}\right.
\label{potential}
\end{eqnarray}
where  $V_1$ is the bulk potential, $a$ is the lattice spacing
in the direction of the surface, $V_0$ is the work function
measured from the bottom of the conduction band, $z_{\rm im}$ is the
image plane, chosen in such a way that the image potential joins
the constant potential $-2V_{1}$ continuously, and is related to
$z_{1}$ as
\begin{equation}
z_{\rm im}=z_{1}-\frac{e^{2}}{4(V_{0}+2V_{1})}.
\end{equation}
In the potential Eq.~(\ref{potential}), $z_{1}$ is the only
free parameter. The eigenfunctions and eigenenergies for the
Schr\"odinger equation associated with the potential (\ref{potential})
can be calculated exactly and the details are given in Appendix A.

\begin{figure}[htb]
\begin{center}
 \includegraphics[width=8cm,keepaspectratio]{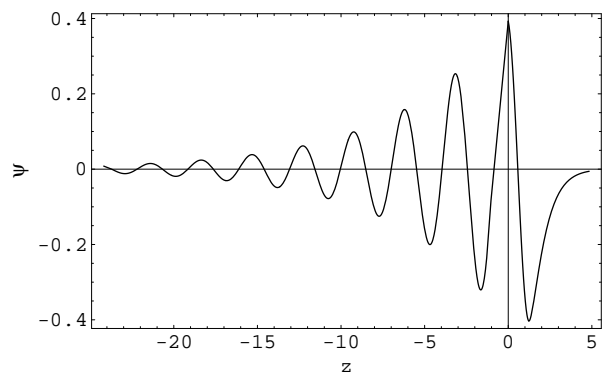}
\end{center}
 \caption{The Cu(111) surface state with energy
$0.39$ eV below the Fermi surface. The vertical axis represents its
normalized wavefunction multiplied by the box size $L$.
The horizontal axis represents $z$ in $\mbox{
A}\!\!\!^{^{^{^{_{_{\circ}}}}}}$.} \label{fig-wavefunction}
\end{figure}

Let us consider first the case of the (111) direction.
In order to fit the
experimental band structure\cite{Knapp} along the (111)
direction in the nearly-free-electron potential
we find that $V_{1}=2.55$ eV, $V_{0}=13.55$ eV.
These values provide a very good description of the band
gap in the (111) direction and also of the ionization energy
of metallic Cu.
The Cu(111) surface has a band of surface states as measured
by photoemission, starting $390$
meV below the Fermi surface\cite{Knapp,Kevan}. By solving the
Schr\"odinger problem for the potential (\ref{potential})
we find the surface band (see Appendix A) and fix the value
of $z_{1}=1.66\mbox{ A}\!\!\!^{^{^{^{_{_{\circ}}}}}}$
in order to fit the photoemission results.
Using this value for $z_{1}$, we
also obtain the next surface state at $788$ meV below the vacuum level.
This result agrees well with photoemission experiments that find the
second surface state at $830$ meV below the vacuum level\cite{Giesen}.
Thus, we were able to get a consistent description of two independent
experiments with a single fitting parameter.  In Fig. \ref{fig-wavefunction}
we show a plot of the surface state wavefunction as a function
of the distance away from the surface (the wavefunction is chosen to
be a real function). Notice the slow decay of
this state inside the crystal in the direction perpendicular to the surface.

In the 3-dimensional $k$-space of Cu, a gaped ``cavity"  is
centered around the (111) direction near the Fermi level and has
the bulk band surrounding this gaped region \cite{Smith}. By
decomposing the momentum of a bulk state $\bf{k}$ into
$\bf{k}_{\perp}$ (perpendicular to the surface) and
$\bf{k}_{\parallel}$ (parallel to the surface), one expects that
the larger $k_{\perp}$, the larger wavefunction tail outside the
crystal, which makes a significant difference on their
contributions to the surface Kondo effect. As a result, it is
important to carefully exclude the (111) gaped region from the Fermi sphere
when calculating the surface Kondo effect.

It is known that a nearly-free-electron band has a parabolic
dispersion not too close to the band edge, where ``band edge"
refers to the interface between the bulk band and the gap in the
3D $k$-space. When calculating the STM tunneling conductance,
one would have to include the contribution
close to the band edge as well.
However, the band-edge contribution is typically much smaller than
the one coming from the 
parabolic band, so we can take merely the contribution from
the parabolic band. For the Cu(111) band, we terminate the
parabolic band $0.7$ eV below the band edge. As long as we work in
a parabolic band, the bulk-state wavefunction inside the crystal
takes a plane-wave form.

We now turn to the case of the (100) direction. We find that
we get good agreement with band
structure calculations \cite{Knapp} by choosing $V_{1}=3.05$ eV,
and $V_{0}=13.45$ eV. In contrast
to the (111) direction, the (100) direction 
has no true surface state near the Fermi
surface. However, there is a surface resonance at $1.15$ eV above
the Fermi level \cite{sr100} that can be well described by choosing
$z_{1}=1.51\mbox{ A}\!\!\!^{^{^{^{_{_{\circ}}}}}}$.
When calculating the tunneling conductance, we can safely
sum over the entire Fermi sphere and adopt a parabolic dispersion,
i.e. a plane-wave wavefunction inside the crystal.

It is the d-orbital of the adatom that actually participates
in the Kondo resonance.
When one of these atoms is placed on a metal surface, its outmost s-wave
electrons either get transfered to the metal conduction band or
to its own d-orbital. For example, the Co adatom tends to form
[Ar]$3d^{9}$ on a Cu surface\cite{Ujsaghy} and therefore has an
effective spin $1/2$. As a result, an adatom d-wavefunction can
be modeled as a Hydrogen atom with an effective charge $Z_{\rm
eff}$ given by:
\begin{equation}
\psi_{a}({\bf r})= R_{n2}(Z_{\rm eff},r)Y_{2m}(\theta,\phi) \; ,
\label{Co}
\end{equation}
where $Y_{2m}(\theta,\phi)$ are Spherical Harmonics and
$R_{n2}$ are Hydrogen radial wavefunctions.
The effective charge $Z_{\rm eff}$ can be determined by comparing
$r_{\rm max}$, defined by,
\begin{equation}
\left.\frac{d|\psi_{a}|^{2}}{dr}\right|_{r=r_{\rm{max}}}=0 \, ,
\end{equation}
with the experimentally observed d-orbital radius 
for metallic Co, $r_{\rm{max}} = 1.25$  \AA. 
In our calculation we orient the axis of quantization
for the atomic  problem  along the direction perpendicular to
the surface.

The STM tip wavefunction is modeled following Tersoff and
Hamann\cite{Tersoff} who proposed a wavefunction of the form:
\begin{eqnarray}
\psi_{p}=\left\{\begin{array}{ll}
  Re^{\kappa R}\frac{\,\,\exp(-\kappa r)\,\,}{r}, & \mbox{for }r>R, \\
  1, & \mbox{for }r<R, \\
\end{array}\right.
\label{tip}
\end{eqnarray}
where $R$ is the radius of curvature of the tip about its center,
the decay constant
$\kappa=\sqrt{2m_{t}^{\ast}(\phi-\epsilon_{p})}$ controls the
wavefunction tail, and $\phi$ is the tip work function.
Notice that the adatom and tip wavefunctions are normalizable,
that is, $\int d^3r |\psi_a({\bf r})|^2 = \int d^3 r |\psi_p({\bf r})|^2 =1$,
when integrated over the entire space.

The set of three different wavefunctions (substrate, adatom, and
tip) are the main elements required for the calculation of the
hybridization energies in the Anderson impurity Hamiltonian.

\section{Hybridization energies}
\label{sec-hybridization}

Using the results of Eq.(\ref{surface-wavefunction}), (\ref{Co})
and (\ref{tip}) we can evaluate the overlap integrals in
(\ref{tunneling}) that are important for the hybridization
energies. Let us first remark that the normalization of each of
these wavefunctions is different; for instance, the bulk state is
normalized in a box of size $L^3$, where $L$ is one of the
dimensions of the box. The surface state, however, since it decays
exponentially away from the surface but is extended on the
surface, is normalized in an area of size $L^2$. Finally, the
adatom and tip states always decay exponentially and therefore do
not require any box normalization. Thus, it is convenient to
redefine the wavefunctions so that:
\begin{eqnarray}
\psi_{B,{\bf k}}({\bf r}) &=& \frac{1}{L^{3/2}} \phi_{3,{\bf k}}({\bf r})
\nonumber
\\
\psi_{S,{\bf k}}({\bf r}) &=& \frac{1}{L} \phi_{2,{\bf k}}({\bf r})
\end{eqnarray}
where the subscript $B$ refers to bulk and $S$ to surface.
Note that in this case the hybridization energies between
substrate (bulk and surface) and adatom
in (\ref{tunneling})
can be written as:
\begin{eqnarray}
t_{{\bf k},a,\alpha} = (\epsilon_{a}+\epsilon({\bf k}))
\frac{1}{L^{d/2}} V_{\alpha}^{(d)}({\bf k})
\end{eqnarray}
where $d=3$ for bulk and $d=2$ for surface. The matrix
elements are, therefore,
\begin{eqnarray}
V^{(d)}_{\alpha}({\bf k}) = \int d^3 r \phi_{d,{\bf k}}({\bf r}) \psi_a({\bf r})  \, .
\label{vd}
\end{eqnarray}
Notice that $V^{(d)}_{\alpha}({\bf k})$ is not only a function of
the momentum and the distance $z_0$ between adatom and surface,
but also depends on which surface - here labeled by the subscript
$\alpha$ - the adatom is located.

\begin{figure}[htb]
\begin{center}
\includegraphics[width=8cm,keepaspectratio]{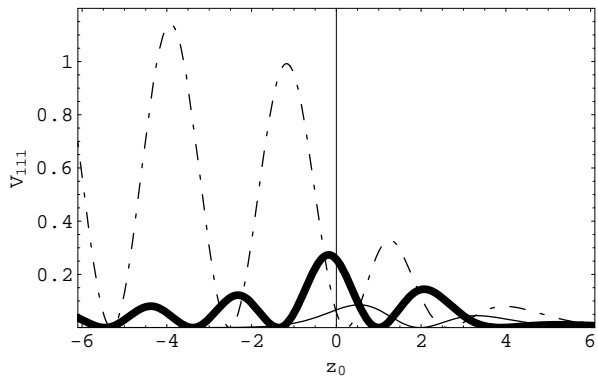}
\end{center} 
\caption{$V_{111}^{(2)}$: dashed-dotted line,
$V_{111}^{(3)}(\Gamma$-$K)$: thick line, and
$V_{111}^{(3)}(\Gamma$-$X)$: thin line, as a function of $z_0$.}
\label{fig-V111}
\end{figure}

In Fig.~\ref{fig-V111} we show the result for $V^{(2)}_{(111)}$,
the hybridization matrix element between $m=0$ adatom state and
surface state in the Cu(111) surface, as obtained by numerical
integration of (\ref{vd}) using (\ref{surface-wavefunction}) and
(\ref{Co}), as a function of $z_0$ (in units of the distance
between the planes in the (111) direction) by assuming ${\bf k}$
to be at the Fermi surface of the surface band [since the surface
states have essentially a spherical Fermi surface there is no
dependence of $V_{111}^{(2)}$ with the direction on the Cu(111)
surface]. The first striking result is the strong oscillation of
the value of the matrix element with $z_0$. This oscillation is
the result of the interference between the d-state of the adatom
and the surface state. Naturally, the largest overlap occurs at
the surface (where the surface state is maximum, see
Fig.~\ref{fig-wavefunction}) but we notice that there is
substantial overlap between the surface state and the adatom at
one lattice spacing from the surface. In principle, we cannot tell
what is the actual orientation of the adatom relative to the
surface but we have checked that the $m=0$ state is the one with
biggest overlap as compared with higher angular momentum states,
thus, instead of a sum over the angular momentum states in
Eq.(\ref{Hamiltonian}) we could have kept only the one with $m=0$. 
Furthermore, first principle calculations
for adatoms in metallic surfaces indicate that the atomic orbitals
are oriented in such a way to generate the largest overlaps \cite{lda_surface}.

In Fig.~\ref{fig-V111} we also plot the value of
$V^{(3)}_{(111)}$, the hybridization matrix element between 
the bulk states and
adatom in the Cu(111) surface as a function of the distance from the
surface for two different bulk states, all of the them at the Cu
Fermi surface, with momenta along the $\Gamma$-X, and $\Gamma$-K
directions (see Fig.\ref{fs}). Once again we notice the oscillations
due to the interference between those states and the adatom states
but we also observe that the overlap now is a factor of 3 smaller
because these states do not have a large amplitude at the surface.
Another interesting result of these calculations is the fact that
as the momenta moves towards the (111) direction the overlap
grows. This is due to the fact that away from the (111) direction
the Fermi surface of Cu is essentially spherical and therefore the
states at the Fermi surface have essentially the same amplitude,
however, the states with large momentum along the direction of the
Cu(111) surface have bigger penetration on the potential
barrier generated by the work function and therefore larger
hybridization with the adatom. This effect will be very important
later for the interpretation of the experimental data.

\begin{figure}[htb]
\begin{center}
\includegraphics[width=8cm,keepaspectratio]{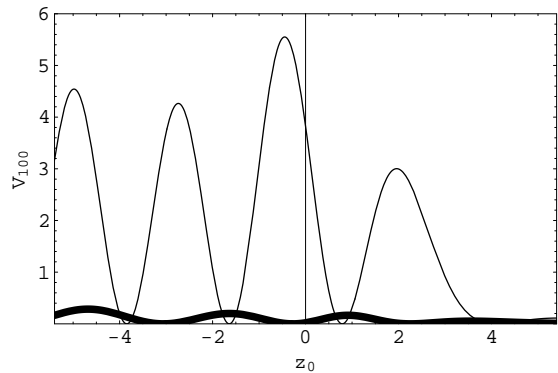}
\end{center}
\caption{$V_{100}^{(3)}(\Gamma$-$X)$: thin line, 
$V_{100}^{(3)}(\Gamma-K)$: thick line, as a
function of $z_0$.} 
\label{fig-V100}
\end{figure}

In Fig.~\ref{fig-V100} we show the value of $V^{(3)}_{(100)}$ for
the overlap between the state of an adatom on the Cu(100) surface
with bulk states with momenta along the $\Gamma$-X (this direction
is perpendicular to the Cu(100) surface), and $\Gamma$-K (since
there are no surface states for the Cu(100) surface we do not need
to calculate $V^{(2)}$ for this surface). Besides the oscillations
found in the previous calculations, we find that when the momentum is
oriented perpendicular to the surface its penetration is maximum
and therefore has largest overlap. In fact, we find that the
overlap between the adatom state on a Cu(111) surface with the
surface state is essentially the same value as the overlap between
the state of the adatom on the Cu(100) surface with a bulk state
with momentum oriented in the (100) direction. One can easily
understand this effect from the nearly free electron picture: in
the absence of the periodic potential of the crystal, the Fermi
surface states in the (111) and (100) direction would be
essentially identical and their decay in the vacuum of their
respective surfaces would be the same. When the periodic potential
of the crystal is added a gap opens in the (111) direction but not
on the (100) direction. However, the bulk state that disappeared at the
Fermi surface in the (111) direction now becomes the surface state
in this same surface. Therefore, its amplitude at the surface and
close to it is essentially the same as the state in the (100)
direction (this is can be readily understood due to the long decay
rate of the surface state shown in Fig.~\ref{fig-wavefunction})
and the overlap is essentially the same. Thus, one has to be
careful when discussing what happens to adatoms in different
surfaces because there is very little difference between surface
states along gaped directions and bulk states along non-gaped
directions in the Fermi surface of noble metals.

\section{The Surface Kondo Effect}
\label{sec-surface-Kondo}

There are two main limits of interest in the Kondo problem. The first one
is reached when $|t_{{\bf k}a}| \ll |\epsilon_a|,U$ and we can
treat the hybridization energy as the smallest scale in the problem.
This is the so-called Kondo regime where the average occupation of the
impurity state is essentially constant, that is, $\langle n_a \rangle \approx 1$.
In order to study this limit it is convenient to split the electron
states into bulk (B) and surface (S) so that we write the Anderson impurity 
Hamiltonian
(\ref{Hamiltonian}) as:
\begin{eqnarray}
H &=& \sum_{{\bf k},\sigma} \epsilon_{1}({\bf k}) c^{\dagger}_{1,{\bf k}\sigma}
c_{1,{\bf k}\sigma} + \sum_{{\bf q}\sigma} \epsilon_2({\bf q})
c^{\dagger}_{2,{\bf q}\sigma} c_{2,{\bf q}\sigma}
\nonumber
\\
&+& \epsilon_a \sum_{\sigma} d^{\dagger}_{\sigma}d_{\sigma}
+ U n_{a,\uparrow}n_{a,\downarrow}
\nonumber
\\
&+& \sum_{{\bf k},\sigma} t_{B,{\bf k}a} \left(c^{\dagger}_{1,{\bf k}\sigma}
d_{\sigma} + h.c.\right)
\nonumber
\\
&+& \sum_{{\bf q}a} t_{S,{\bf q}a}
\left(c^{\dagger}_{2,{\bf q}\sigma} d_{\sigma} + h.c.\right)
\label{two_band_anderson}
\end{eqnarray}
where $c_{1,{\bf k}\sigma}$ ($c_{2,{\bf q}\sigma}$) describes the bulk 
(surface)
electron annihilation operator, respectively. We have a two-band problem
where both bulk and surface states hybridize with the impurity at the
surface. Notice that this is a single channel problem since only
one effective channel couples to the impurity.

In the Kondo limit we can use a Schrieffer-Wolf transformation
\cite{schrieffer_wolf} and rewrite the Anderson impurity Hamiltonian as a
exchange problem:
\begin{eqnarray}
H &=& \sum_{{\bf k},\sigma,n=1,2} \epsilon_{n}({\bf k})
c^{\dagger}_{n,{\bf k}\sigma} c_{n,{\bf k}\sigma}
\nonumber
\\
&+& \sum_{{\bf k},{\bf k'}} \sum_{\alpha,\beta=\uparrow,\downarrow}
    \sum_{n,m=1,2} {\bf S} \cdot \vec{\sigma}_{\alpha,\beta}
c^{\dagger}_{n,{\bf k},\alpha} J_{n,m}({\bf k},{\bf k'}) c_{n,{\bf k},\beta}
\label{Kondo_Hamiltonian_2}
\end{eqnarray}
where ${\bf S}$ is the localized impurity spin and
\begin{eqnarray}
J_{n,m}({\bf k},{\bf k'}) = \frac{1}{L^{(d_n+d_m)/2}}
\frac{V^{(d_n)}({\bf k}) V^{(d_m)}({\bf k'})}{\lambda}
\label{exchanges}
\end{eqnarray}
is the exchange energy matrix (we have defined $d_1 = 3$ and $d_2=2$ as the
dimensionalities of bulk and surface states, respectively), and
\begin{eqnarray}
\frac{1}{\lambda}=\frac{1}{|\epsilon_a|}
+\frac{1}{|\epsilon_a+U|}\, .
\end{eqnarray}
Notice that, unlike the single band Kondo effect (\ref{Kondo_Hamiltonian_1}),
electrons from one band can scatter from the impurity between two different bands with
spin flip (all energies are measured relative to the Fermi energy).
If we ignore the off-diagonal matrix elements of (\ref{exchanges})
the problem trivially
reduces to two independent Kondo effects with characteristic Kondo
temperatures
\begin{eqnarray}
T_{K,n} \approx D_n e^{-1/g_n}
\label{tk1}
\end{eqnarray}
where $D_n$ is the bandwidth cut-off ($D_1 \approx 8.6$ eV, $D_2 \approx 0.4$
eV \cite{Knapp}) and
\begin{eqnarray}
g_n = N_n(0) J_{n,n}(k_F,k_F) = \frac{1}{L^{d_n}}
\frac{N_n(0)|V^{(d_n)}(k_F)|^2}{\lambda} \label{ln}
\end{eqnarray}
is the effective coupling constant, $N_n(0)$ is the Fermi surface
density of states (for spherical Fermi surface we have $N_n(0) =
(L/\pi)^{d_n} 2 \pi m^* k_F^{d_n-2}/\hbar^2$ where $m^*$ is the
effective mass). Because of the difference in the Kondo
temperatures given in (\ref{tk1}) one of the scattering channels dominates
the screening process and the singlet state can be either formed
with the surface state when $T_{K,2}>T_{K,1}$ or with the bulk
states if $T_{K,1}>T_{K,2}$. If we apply this argument to the
results of the overlap integrals of the previous section we see
that for the adatom in the Cu(111) surface the Kondo screening is
dominated by the surface state since the coupling constant
(\ref{ln}) is bigger for this state than the bulk states. On the
other hand, if the adatom is sitting at the Cu(100) surface, the
Kondo effect is dominated by the bulk state with momentum
perpendicular to the Cu(100) surface.

The above discussion is valid as far as the occupation of the
impurity level is close to $1$. However, in the problem at hand
the adatom occupation can change substantially with the
hybridization since the matrix element is strongly dependent of
the distance to the surface and the type of states involved, that
is, $t_{{\bf k}a}$ can be of the order of $\epsilon_a$ (but still
much smaller than $U$). In this case, the mapping into the Kondo
problem fails to properly describe the actual occupation on the
adatom. Instead, we should start from the Anderson impurity Hamiltonian
(\ref{Hamiltonian}) and compute the Kondo temperature directly
from it. Here we will do so by adopting the large $N$ calculation
of Newns and Read \cite{Newns&Read} that allows for average
fluctuations in the valence of the adatom in the limit of $U \to
\infty$ (in fact, estimates based on LSDA\cite{Ujsaghy} give a
value of $U\approx2.84$ eV and $\epsilon_{a}\approx-0.84$ eV). The
idea of the large $N$ calculation is to generalize
(\ref{Hamiltonian}) to include more degenerate electron states
(besides the $2$ states associated with the spin $1/2$) and take
the limit where the degeneracy diverges. Mathematically this is
accomplished by generalizing the spin index in the electron
operator to a new index $m$ that varies from $1$ to $N$. It is
convenient to write Anderson impurity Hamiltonian (\ref{Hamiltonian}) in
the limit of $U\to\infty$ as:
\begin{eqnarray}
H_{s}&=&\epsilon_{a}\sum_{m} c_{a,m}^{\dagger}c_{a,m}
+\sum_{{\bf k},m}\epsilon_{k}c_{{\bf k},m}^{\dagger}c_{{\bf k},m}
\nonumber \\
&&+\sum_{{\bf k},m}t_{{\bf k},a}\left(c_{{\bf k}m}^{\dagger}c_{am} {\cal P}
+{\rm{H}}.{\rm{c}}.\right)
\label{largeN}
\end{eqnarray}
where ${\cal P}$ is the projector operator that projects out any
states of the adatom with double occupancy. In the adatom we can,
in principle, have three different states with zero, $|d^0\rangle$,
single, $|d^1 \rangle$, and double, $|d^2 \rangle$, occupancy. We now
expand the Hilbert space by introducing a ``phantom'' boson state such
that when the state of the adatom is unoccupied by an electron it
is occupied by a boson, say $|b^1 d^0\rangle$, and when it is unoccupied
by a boson it is occupied by an electron, $|b^0 d^1\rangle$. Observe
that this construction constrains the number of bosons and electrons
to be $1$ for all states $|\psi\rangle$ in the Hilbert state:
\begin{eqnarray}
(n_a + b^{\dagger} b)|\psi\rangle = 1 |\psi\rangle \, .
\label{constraint}
\end{eqnarray}
The utility of this representation becomes clear when we consider
the action of the operator $c^{\dagger}_{am} b$ on the state $|b^1
d^0\rangle$ producing the state $|b^0 d^1\rangle$. Further
application of $c^{\dagger}_{am} b$ to $|b^0 d^1\rangle$
annihilates the state while application of $c_{am} b^{\dagger}$ to
$|b^0 d^1\rangle$ returns the system to the state $|b^1
d^0\rangle$. Thus, the repetitive application of $c^{\dagger}_{am}
b$ or $c_{am} b^{\dagger}$ never produces the doubly occupied
state. In other words, we can replace ${\cal  P}$ in
(\ref{largeN}) by a boson operator if we ensure that the
constraint (\ref{constraint}) is obeyed. This constraint can be
seen as an energy cost, that is, states of the Hamiltonian that
violate the constraint (\ref{constraint}) are highly excited states. 
If we are interested in the low energy physics then
we can convert (\ref{constraint}) into a new energy scale and add
it to the Hamiltonian as:
\begin{eqnarray}
H_{s}&=&\epsilon_{a}\sum_{m} c_{a,m}^{\dagger}c_{a,m}
+\sum_{{\bf k},m}\epsilon_{k}c_{{\bf k},m}^{\dagger}c_{{\bf k},m}
\nonumber \\
&&+\sum_{{\bf k},m}t_{{\bf k},a}\left(c_{{\bf k}m}^{\dagger}c_{am} b
+{\rm{H}}.{\rm{c}}.\right)
\nonumber \\
&+& \gamma (n_a + b^{\dagger} b -1)
\label{largeN_1}
\end{eqnarray}
where $\gamma$ is a Lagrange multiplier. Notice that in the limit
of $\gamma \to \infty$ the constraint (\ref{constraint}) is automatically
obeyed since the energy is minimized when (\ref{constraint}) is fulfilled.
As is, Hamiltonian (\ref{largeN_1}) is still very complicated and
at this point we have to use the large $N$ approximation that consists
in treating (\ref{largeN_1}) in a mean-field approximation. In the mean-field
limit we consider the average value of the boson fields instead of their
actual value, that is, we replace the operators $b$ and $b^{\dagger}$ by
\begin{eqnarray}
\langle b \rangle = \langle b^{\dagger} \rangle = \sqrt{z} \, .
\label{mf_bosons}
\end{eqnarray}
Notice that by treating the bosons in the mean-field approximation we
obtain a quadratic Hamiltonian in (\ref{largeN_1}) that can be diagonalized
by standard techniques and the total energy of the system can be minimized
with respect to $z$ and $\gamma$. The physical meaning of these parameters
becomes clear when we calculate physical observables. For instance,
the average occupation of the impurity is given by:
\begin{eqnarray}
\langle n_a \rangle = 1 - z
\end{eqnarray}
that shows that $z$ has to do with the change in the occupancy of the
adatom state due to the hybridization with the substrate. On the other
hand we can also show that $\gamma$ is the renormalization of the energy
of the impurity state, that is, the renormalized impurity state energy
is given by: $\bar{\epsilon}_a = \epsilon_a + \gamma$.
The mean-field parameters $\bar{\epsilon}_{a}$ and $z$ are
calculated by solving a set of coupled transcendental equations\cite{Newns&Read}:
\begin{eqnarray}
1-z&=&\frac{2}{\pi}\arctan\left(\frac{zF_{aka}(\epsilon_{F})}{|\bar{\epsilon}_{a}|}\right),
\nonumber \\
\epsilon_{a}-\bar{\epsilon}_{a}&=&\frac{2}{\pi}F_{aka}(\epsilon_{F})\ln\left(
\frac{\left\{\bar{\epsilon}_{a}^{2}+\left[zF_{aka}(\epsilon_{F})\right]^{2}\right\}^{1/2}}
{D}\right),
\end{eqnarray}
where $F_{\alpha k\beta}(\omega)$ are matrices defined, in general, as:
\begin{equation}
F_{\alpha{k}\beta}(\omega)=\sum_{{\bf k}}
\frac{t_{\alpha{{\bf k}}}t_{{\bf k}\beta}}{\omega-\epsilon_{{\bf k}}+i\eta},
\label{F-Green}
\end{equation}
and $D$ is bandwidth of the conduction band (notice that
$\alpha,\beta = a,p$ in reference to the adatom and tip, respectively).
The Kondo temperature can be directly extracted from this formalism as \cite{Newns&Read}
\begin{eqnarray}
k_{B}T_{K}=z\Delta_{0}=D\exp\left(\frac{\pi\epsilon_{a}}{\Delta_{0}}\right).
\label{tk2}
\end{eqnarray}
where
\begin{equation}
\Delta_{0}=\,{\rm{Im}}\,F_{aka}(\omega).
\end{equation}
Therefore, using the large $N$ account we can deal with the case
where the hybridization energies are of the order of the adatom
energies. We use these results in studying the STM differential
conductance in the next section.

\begin{figure}[htb]
\begin{center}
\includegraphics[width=8cm,keepaspectratio]{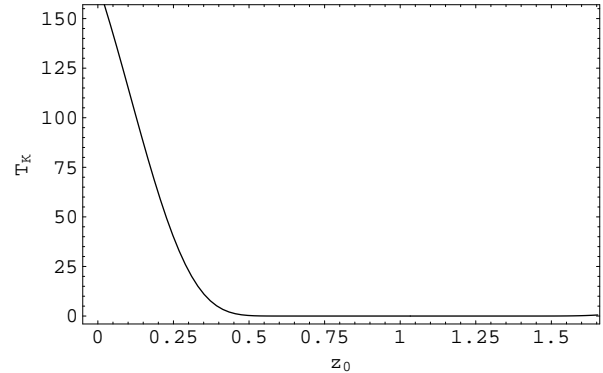}
\end{center}
\caption{Kondo temperatures as functions of the Co
distance from the surface, $z_0$, on the Cu(111) surface.} 
\label{fig-Kondo-T111}
\end{figure}

\begin{figure}[htb]
\begin{center}
\includegraphics[width=8cm,keepaspectratio]{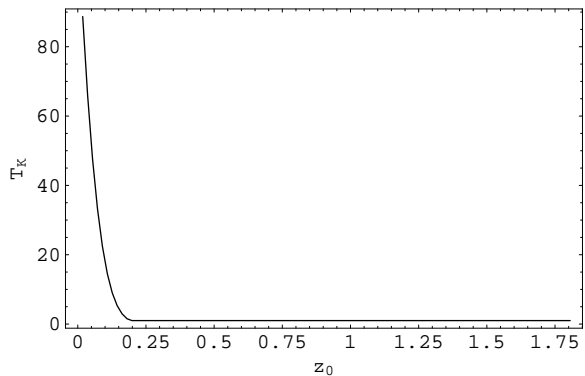}
\end{center}
\caption{Kondo temperature as functions of the Co
distance, $z_0$, on the Cu(100) surface.} 
\label{fig-Kondo-T100}
\end{figure}

We are not aware of any measurement of Co-to-Cu distance being
reported, so this distance $z_{0}$ will be treated as a parameter
of the order of the Cu lattice space. Fig.~\ref{fig-Kondo-T111}
and \ref{fig-Kondo-T100} shows our calculated Kondo temperature as
a function of $z_{0}$. An interesting feature is that the Kondo
temperature decays fast with $z_{0}$, which can be understood because
of the exponential overlap between the substrate and Co wavefunctions. When
calculating the Kondo temperature for Co/Cu(111) assuming that
only the bulk state exists, we find $T_{K}<10^{-6}K$ for the
$z_{0}$'s as Fig.~\ref{fig-Kondo-T111}, which clearly means that
the surface state dominates the Kondo effect. The domination is
not surprising since a significant portion of large-$k_{\perp}$
bulk states are gaped around the (111) direction. Comparing the
Cu Fermi surface with a perfectly spherical one, one finds that
the Cu surface state contributes to the Kondo effect in a way
similar to the bulk band of the spherical Fermi surface around the
(111) direction. Similarly, the bulk band with respect to the
Cu(100) surface can be considered as a perfectly spherical Fermi
surface for calculating the Kondo effect. As a result, the Kondo
effect of the Co/Cu(111), dominated by the surface state, should
have the same order of magnitude as Co/Cu(100). The positions that
provide the experimentally measured Kondo temperatures \cite{Knorr}
($T_{K,(111)}=54$K and $T_{K,(100)}={88}$K) are $0.41\mbox{
A}\!\!\!^{^{^{^{_{_{\circ}}}}}}$ and $0$ from the (111) and (100)
surfaces, respectively. Hence the STM conductance will be
calculated by fixing Co atom at these optimized positions.

We should point out that from an experimental point of view 
we would expect the
Co adatom to be located around one lattice spacing away from
the surface. Thus, our obtained values of their position seem
to be quite small (especially in the case of the (100) surface
where the adatom has to be located at the surface) and therefore
unrealistic. Nevertheless, one has to have in mind that the
there are many assumptions made in the calculation that can 
affect significantly the value of the
Kondo temperature, namely, the form of the image potential close to 
the surface, the large $N$ approximation, and the limit of
$U \to \infty$. Furthermore, it is clear from (\ref{tk2}),
that the Kondo temperature is exponentially sensitive on the
details of the hybridization and energy scales and therefore
is significantly affected by the various assumptions. Thus,
the plots in Fig.\ref{fig-Kondo-T111} and Fig.\ref{fig-Kondo-T100}
have to taken as the trends for the variation of the Kondo
temperature with the distance from the surface and not
as realistic values.

\section{STM Differential Conductance}
\label{sec-STM}

We assume that the STM tip and the adsorbate-substrate complex are
each in local equilibrium, and use the standard Green's function 
formalism for
calculation of the differential conductance \cite{Gadzuk}:
\begin{equation}
\frac{dI}{dV}\propto-{\rm{Im}}\,\left\{F_{pkp}(\omega)+G_{a}(\omega)
\left[t_{pa}+F_{pka}(\omega)\right]^{2}\right\}_{\omega=eV},
\label{current}
\end{equation}
where $F_{\alpha k\beta}(\omega)$ is defined in (\ref{F-Green})
and $G_{a}$ is formally the adsorbate retarded Green's function
associated with the total Hamiltonian $H$. However, because the
effect on the adsorbate from the tip is much smaller than that
from the substrate, it is always safe to regard $G_{a}$ as the one
without tip. In this work, $G_{a}$ (without tip) is calculated by
the mean-field slave-boson technique \cite{Newns&Read} as:
\begin{equation}
G_{a}(\omega)=\frac{1}{\omega-\bar{\epsilon}_{a}+iz\Delta_{0}},
\label{ga}
\end{equation}
where the symbols here are given in the previous section.

By substituting (\ref{ga}) into (\ref{current}), the STM differential
conductance can be written as:
\begin{equation}
\frac{dI}{dV}-\left(\frac{dI}{dV}\right)_{0} = A
\left[\frac{q^{2}-1+2q\xi}{\xi^{2}+1}\right] \, ,
\label{didv}
\end{equation}
where the differential conductance with a subscript ``0" refers to
the background signal, $A$ is the amplitude of the STM conductance,
\begin{eqnarray}
A \propto \left({\rm{Im}}F_{pka}\right)^2 (1+q^2) \, \, ,
\label{amp}
\end{eqnarray}
$q$ is the Fano lineshape parameter, 
\begin{equation}
q=\frac{t_{pa}+{\rm{Re}}\,F_{pka}(\omega)}{{\rm{Im}}\,F_{pka}(\omega)},
\label{Fano}
\end{equation}
and $\xi$ is the dimensionless bias voltage
\begin{equation}
\xi=\frac{eV+\bar{\epsilon}_{a}}{k_{B}T_{K}}.
\end{equation}
Expression (\ref{didv}) has been frequently used for the
fitting of STM data. 

\begin{figure}[htb]
\begin{center}
 \includegraphics[width=8cm,keepaspectratio]{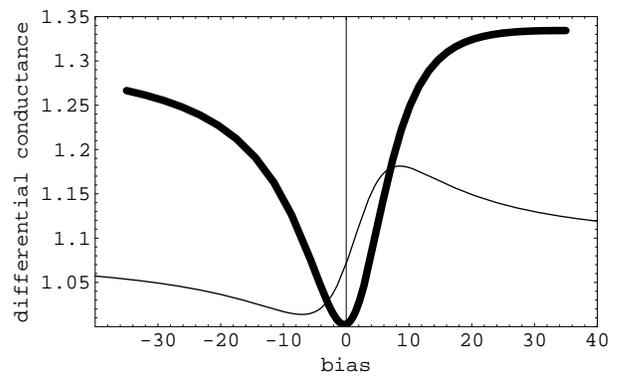}
\end{center}
 \caption{Differential conductance (in arbitrary units) 
at $R_{\parallel}=0$, as a
function of the bias $V$ in meV for an adatom on the Cu(111) 
(thick line) and Cu(100) (thin line) surfaces, as obtained
from the theoretical calculations.}
\label{dIdV} 
\end{figure}

\begin{figure}[htb]
\begin{center}
 \includegraphics[width=8cm,keepaspectratio]{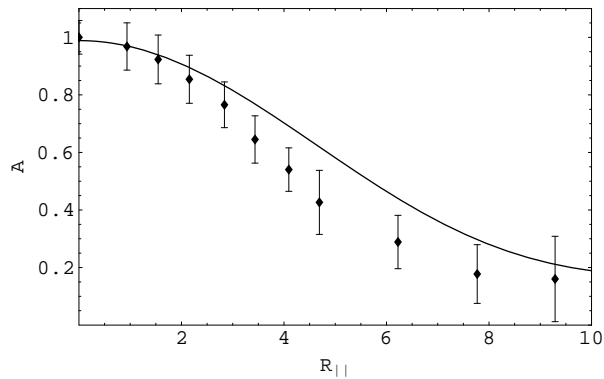}
\end{center}
 \caption{Amplitude of the differential conductance (in arbitrary units) 
for Co on Cu(111) 
as a function of $R_{\parallel}$ as compared with the experimental
data \cite{Knorr}.}
\label{amplitude} 
\end{figure}

The shape of the differential conductance versus the bias voltage is
determined by the Fano line shape parameter $q$. 
In Fig.\ref{dIdV} we show the differential conductances obtained
from our microscopic theory for the Cu(111) and Cu(100) surfaces. 
From Eq.~(\ref{Fano}) we see that $q$ is related to three microscopic
quantities $t_{pa}$, ${\rm{Im}}\,F_{pka}(\omega)$, and
${\rm{Re}}\,F_{pka}(\omega)$. It is found that $t_{pa} \ll
{\rm{Re}}\,F_{pka}(\omega)$ and therefore we can set $t_{pa}=0$ 
in (\ref{Fano}).
Obtaining the first two is
straightforward since they only depend on
the physics {\it at} the Fermi energy. From (\ref{F-Green}), we
have:
\begin{eqnarray}
{\rm{Im}}\,F_{\alpha k \beta}(\omega) = \pi\sum_{{\bf k}}
t_{\alpha{\bf k}} t_{{\bf k}\beta} \delta(\omega-\epsilon_k) \, .
\end{eqnarray}
Notice that the amplitude of the STM signal, (\ref{amp}), 
depends essentially on this quantity and can be calculated
reliably. In Fig.\ref{amplitude} we plot $A(R_{\parallel})$
for $R_{\perp} = 6$ \AA and compare it with the experimental 
data \cite{Knorr}. As
one can see, the agreement between the theory and experiment
and theory is quite good.

\begin{figure}[htb]
\begin{center}
 \includegraphics[width=8cm,keepaspectratio]{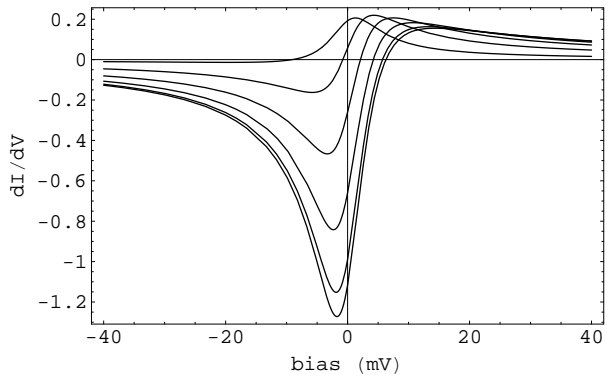}
\end{center}
 \caption{Differential conductances for the Co on Cu(111) for
different values of $R_{\parallel}$. From bottom to the top:
$R_{\parallel}=0,2,4,6,8,10$ \AA} 
\label{didv_r}
\end{figure}

The other quantity of interest, namely, ${\rm{Re}}\,F_{pka}(\omega)$, 
requires an integration over the entire Cu band. For the Cu(111) surface, the
surface band has a free-electron dispersion starting $390$ meV
below the Fermi level up to $400$ meV above it, where the upper
limit is the intercept of the surface band to the bulk bands. We
calculate ${\rm{Re}}\,F_{pka}(\omega)$ for this surface band and
determine the tunneling conductances, (\ref{didv}), for different 
values of $R_{\parallel}$ as shown in shown in Fig.~\ref{didv_r}.

We notice that for short distances away from the adatom, the theory
can describe very well the lineshape of the 
tunneling conductance. Nevertheless,
as the distance of the STM tip to the adatom is increased, the 
lineshape becomes more symmetric and a positive Lorentzian is
developed (see Fig.\ref{didv_r}). The origin of this result can be easily
understood: notice that in Fig.\ref{amplitude} the amplitude
of the STM conductance becomes very small for large values 
of $R_{\parallel}$. This smallness can be tracked down to the
small value of ${\rm Im}F_{pka}$ in (\ref{amp}) indicating
that the interference between tip and adatom is fading away.
However, one sees from (\ref{Fano}) that this indicates
that the Fano parameter $q$ is becoming very large and positive
given rise to a positive Lorentzian. This feature is general
of any theory of the STM conductance that starts with nearly
free electron wavefunctions and cannot be avoided since it
has its origins on the oscillating nature of the surface states
density of states \cite{Gadzuk}. This result is
at odds with the experiment when the distance between tip and
adatom becomes large . Instead of a positive Lorentzian, it is
observed a negative Lorentzian \cite{manoharan}.

\begin{figure}[htb]
\begin{center}
 \includegraphics[width=8cm,keepaspectratio]{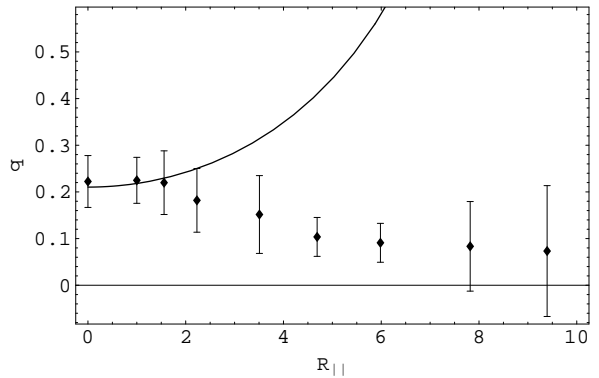}
\end{center}
 \caption{Plot of the Fano parameter $q(R_{\parallel})$ 
for Co on Cu(111) as compared with the values obtained by 
the experimental fit \cite{Knorr}.}
 \label{fig-q}
\end{figure}

In Fig.\ref{fig-q} we plot the calculated
values of $q$ against the values used by the experimentalists
in order to fit the data \cite{Knorr}. The discrepancy between
theory and experiment is clear. More work will be required in 
order to check this issue.

In order to calculate ${\rm{Re}}\,F_{pka}(\omega)$ for an adatom
located at the Cu(100) surface we need information of the whole
Cu bulk band. The Cu bulk bands start at approximately 
$10$ eV below the Fermi level and disperse up to the vacuum
level that is located at approximately at $4.8$ eV above the
Fermi energy. At the Fermi energy the Cu band has dominant
s-wave symmetry which gives the quasi-spherical shape of the
Fermi sea. However, there are also Cu d-bands 
located from $2$ to $5.5$ eV below the Fermi level with
very large density of states \cite{paper_cu}. 
Using the nearly free electron approximation (s-wave) with a cut-off at
the vacuum level we have calculated ${\rm{Re}}\,F_{pka}(\omega)$
and have found that it  
yields a value of $q(R_{\parallel})$ that changes sign
from positive to negative as $R_{\parallel}$ increases, which is
not the case observed in the experiment. In principle the quantity
${\rm{Re}}\,F_{pka}(\omega)$ must be calculated
including the d-bands. Using a simple parabolic dispersion
for the d-bands we find their contribution to be less than 0.1\%
of the total value of ${\rm{Re}}\,F_{pka}(\omega)$ and therefore
these bands are not able to fix the problem of the sign of $q(R_{\parallel})$. 
Thus, we have problems
with the use of bulk states in our calculation. We believe that
band structure effects are important for the exact calculation of
the surface Kondo effect due to bulk states. As far as we know
there are no such detailed calculations. 

\begin{figure}[htb]
\begin{center}
 \includegraphics[width=8cm,keepaspectratio]{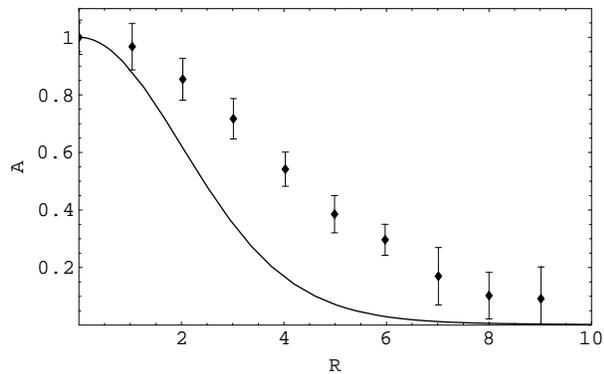}
\end{center}
 \caption{Amplitude of the differential conductance (in arbitrary units) 
for Co on Cu(100) 
as a function of $R_{\parallel}$ as compared with the experimental
data \cite{Knorr}.}
\label{a100} 
\end{figure}

In order to remedy this
problem we introduce a cut-off in the problem such that by
integrating bulk states starting from the bottom of the nearly
free electron band we stop the integration when the value of
${\rm{Re}}\,F_{pka}(\omega)$ fits the experimental curve at
$R_{\parallel}=0$. 
We find that if we integrate up to $0.25$ eV above the Fermi level we can
quantitatively explain the data, as seen in Fig.~\ref{dIdV}.
Given the limitations of our theory this is essentially 
equivalent to  consider ${\rm{Re}}\,F_{pka}(\omega)$
for adatom on the Cu(100) surface for bulk states as a free
parameter of the problem. More accurate calculations, beyond the
ones presented in this paper, are required in order to 
calculate ${\rm{Re}}\,F_{pka}(\omega)$ from a microscopic theory.

\begin{figure}[htb]
\begin{center}
 \includegraphics[width=8cm,keepaspectratio]{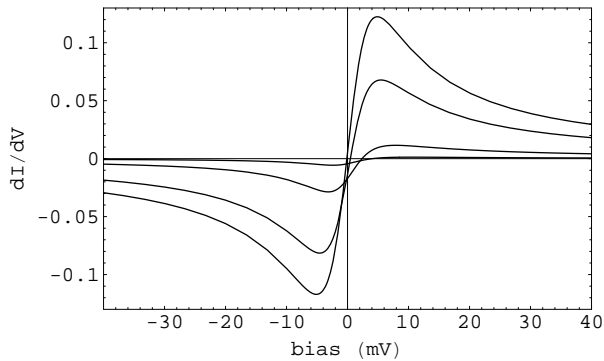}
\end{center}
 \caption{Differential conductances for the Co on Cu(100) for
different values of $R_{\parallel}$. From bottom to the top:
$R_{\parallel}=0,2,4,6$ \AA} 
\label{dIdV_100}
\end{figure}

\begin{figure}[htb]
\begin{center}
 \includegraphics[width=8cm,keepaspectratio]{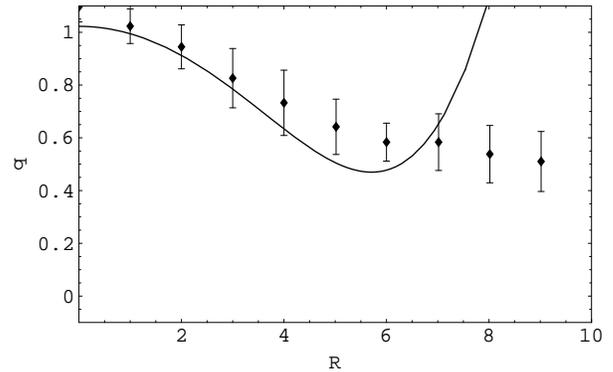}
\end{center}
 \caption{Plot of the Fano parameter $q(R_{\parallel})$ 
for Co on Cu(100) as compared with the values obtained by 
the experimental fit \cite{Knorr}.}
 \label{q100}
\end{figure}

In Fig.~\ref{a100} we show the amplitude $A(R_{\parallel})$ for
Co on Cu(100) for $R_{\perp} = 9$ \AA.  There is a qualitative
agreement between theory and the experiment. 
In Fig.~\ref{dIdV_100} we show the evolution of the tunneling
conductance as a function of the distance from the adatom. 
Once again, the lineshape is correct for short distances but for
long distances the lineshape becomes a positive lorentzian, in
contrast with the experiment. In fact, in Fig.~\ref{q100} we
show the dependence of the Fano parameter, $q$, as a function
of the lateral distance, $R_{\parallel}$. Notice that overall
result is better than the case of Co on Cu(111) (Fig.~\ref{fig-q})
but the divergence in the value of $q$ is very clear. 

\section{Conclusions}
\label{sec-conclusions}

In this paper we have studied the microscopic theory of the single
impurity surface Kondo resonance starting from an 
Anderson impurity Hamiltonian. We have calculated the hybridization
energies and wavefunctions of the model
adopting simple models for the crystal and surface potentials in
the nearly-free electron approximation. Our model reproduces the
main features of the band structure of Cu, the energy of the
surface states and surface resonances as measured in photoemission
experiments.

We have shown that the hybridization energies (or matrix
elements), calculated from the solution of the single particle Schr\"odinger
equation,  are strongly dependent on the direction of the electron momenta,
the distance of the adatom from the surface, and the actual surface where
the magnetic impurity is located. The hybridization matrix elements
oscillate with distance from the surface due to the interference between
the adatom localized state and the substrate states (surface and bulk)
that oscillate close to the surface. We have shown that the surface Kondo
effect occurs preferentially with the surface state (when it exists) or
with the bulk state at the Fermi energy that has the largest component
of its momentum perpendicular to the surface.

We have demonstrated that although the surface Kondo effect can be
obtained from the Anderson impurity Hamiltonian by a Schrieffer-Wolf
transformation onto the Kondo Hamiltonian, the hybridization
energies can be comparable with the adatom atomic energy requiring
a more sophisticated approach that allows for fluctuations in the
mean occupation of the adatom (in other words, the adatom is in
what is called a mixed-valence state). We have used the mean-field
slave-boson formalism that allows for the calculation of the Kondo
temperature in the limit when the on-site Coulomb energy $U$ is
much larger than the hybridization energy. Our results produce
reasonable values of the Kondo temperature for these systems and
the distance of the adatom from the surface.

We have calculated the tunneling conductance of a STM tip for
different surfaces in the presence of a Co impurity and found that
for the surface state in the Cu(111) surface we get quantitative
agreement without any further adjustable parameters. However, for
the adatom on the Cu(100) surface where the Kondo effect is
dominated by the bulk state with momentum along the (100)
direction, we have to add an extra cut-off energy in order to fit
the data. More detailed band structure calculations are required
in order to eliminate the need for this extra parameter. We have
shown that in both cases we can fit very well the data obtained in
STM measurements at $R_{\parallel}=0$ but there is a discrepancy
between the theory and experiments when the distance between the
tip and the adatom becomes large. While in the experimental paper where the
measurements are reported it is conjectured that the Kondo effect
in the Cu(111) and Cu(100) surfaces are dominated by the bulk
states\cite{Knorr}, we show that in fact, the Kondo effect is
dominated by the surface state in the Cu(111) surface and by the
bulk state with momentum in the (100) direction in the (100)
surface. We show that this occurs because there is very little
difference between surface states and bulk states close to the
surface in a noble metal and both states have maximum penetration
on the work function potential.

In summary, we provide a microscopic interpretation for the STM experiments
of surface Kondo effects. Our results can be used as basis for the interpretation
of more complicated situations where, for instance, Co adatoms interact
via direct or indirect exchange on artificial Anderson lattices in metallic
surfaces.

{\bf Acknowledgments} We would like to thank I.~Affleck, P.~W.~Anderson, 
A.~Balatsky, P.~Coleman, D.~Eigler, A.~Heinrich, 
P.~Johnson, C.~Lutz, V.~Madhavan, P.~Simon, and A.~Tsvelik, 
for illuminating discussions. We also
would like to acknowledge support under DARPA contract no. DAAD19-01-C-0060.

\appendix
\section{Explicit Copper Wavefunctions}
\label{appnd}

For the Cu(111) surface state, the parallel momentum
${\bf{k}}_{\parallel}$ is a good quantum number, and the
wavefunction can be solved from potential Eq.~(\ref{potential}) as:
\begin{eqnarray}
&&\psi({\bf{r}})=e^{i{\bf{k}}_{\parallel}\cdot{\bf{r}}_{\parallel}}
 \nonumber \\
&&\times\left\{\begin{array}{ll}
e^{\mu z}\cos(k_{z}z+\delta), & \mbox{for }z<0, \\
{\cal A}\cos\lambda z+{\cal B}\sin\lambda z,
& \mbox{for }0<z<z_{1}, \\
{\cal{C}}z'e^{-\kappa z'}U\left(1-\frac{a}{4\pi
a_{0}\kappa},2;2\kappa z'\right), & \mbox{for }z>z_{1}.
\end{array}\right. \nonumber \\ \label{surface-wavefunction}
\end{eqnarray}
where $U(a,b;x)$ is the standard 2nd solution of
confluent hypergeometric equation\cite{Arfken},
$z'=z-z_{\rm{im}}$, and $k_{z}=\pi/2$. Inside the crystal one 
has a phase shift $\delta$, a crystal decay factor $\mu$, and the
energy of the surface band bottom $E$ related by
\begin{eqnarray}
V_{1}\cos2\delta&=&-\frac{V_{1}^{2}ma^{2}}{2\hbar^{2}\pi^{2}}\sin^{2}2\delta
+\frac{\hbar^{2}\pi^{2}}{2ma^{2}}-E, \nonumber \\
\frac{\hbar^{2}\mu^{2}}{2m}&=&\left(V_{1}^{2}+\frac{2\hbar^{2}\pi^{2}}{ma^{2}}E\right)^{1/2}
-\left(\frac{\hbar^{2}\pi^{2}}{2ma^{2}}+E\right),
\end{eqnarray}
where $m=0.84m_{e}$ is the bulk-band effective mass. 
The plane wavefunction has a wavevector
\begin{equation}
\lambda=\frac{1}{\hbar}\left[2m\left(E+2V_{1}\right)\right]^{1/2} \, ,
\end{equation}
while vacuum decay factor $\kappa$ is defined as
\begin{equation}
\kappa=\frac{1}{\hbar}\left[2m\left(V_{0}-E\right)\right]^{1/2} \, .
\end{equation}
The energy $E$ is obtained from the
continuity conditions at $z=0$ and $z=z_{1}$.


\begin{thebibliography}{90}
\bibitem{kfr}As pointed out in ref.\cite{tsvelik}, the radius of the
d-orbital, $r_B$, is much smaller than the average interparticle distance,
$k_F^{-1}$ (where $k_F$ is the Fermi momentum): $k_F r_B \ll 1$. Hence,
the corresponding overlap integrals between the adatom and the substrate
are small.
\end{thebibliography}
\end{document}